\documentclass[runningheads]{llncs}
\usepackage[T1]{fontenc}
\usepackage{graphicx}
\usepackage{textcomp}
\usepackage{xcolor}
\usepackage{url}
\usepackage{booktabs}
\usepackage[para]{footmisc}
\usepackage{enumitem,kantlipsum}
\usepackage{algorithm}
\usepackage[noend]{algpseudocode}
\usepackage{float}
\usepackage{marvosym}

\newcommand{\algorithmicsize}{\tiny}

\begin{document}
\title{Towards Secure Management of Edge-Cloud IoT Microservices using Policy as Code}
\titlerunning{Secure Microservice Management}

\author{Samodha Pallewatta \inst{1, 2}\textsuperscript{(\Letter)} \and
Muhammad Ali Babar \inst{1, 2, 3} }
\authorrunning{S. Pallewatta and M.A. Babar}
\institute{CREST - The Centre for Research on Engineering Software Technologies
\and
The University of Adelaide, Australia
\and
Cyber Security Cooperative Research Centre (CSCRC), Australia
\email{samodha.pallewatta@adelaide.edu.au, ali.babar@adelaide.edu.au}\\
}

\maketitle              
\begin{abstract}
IoT application providers increasingly use MicroService Architecture (MSA) to develop applications that convert IoT data into valuable information. The independently deployable and scalable nature of microservices enables dynamic utilization of edge and cloud resources provided by various service providers, thus improving performance. However, IoT data security should be ensured during multi-domain data processing and transmission among distributed and dynamically composed microservices. The ability to implement granular security controls at the microservices level has the potential to solve this. To this end, edge-cloud environments require intricate and scalable security frameworks that operate across multi-domain environments to enforce various security policies during the management of microservices (i.e., initial placement, scaling, migration, and dynamic composition), considering the sensitivity of the IoT data. To address the lack of such a framework, we propose an architectural framework that uses Policy-as-Code to ensure secure microservice management within multi-domain edge-cloud environments. The proposed framework contains a "control plane" to intelligently and dynamically utilise and configure cloud-native (i.e., container orchestrators and service mesh) technologies to enforce security policies. We implement a prototype of the proposed framework using open-source cloud-native technologies such as Docker, Kubernetes, Istio, and Open Policy Agent to validate the framework. Evaluations verify our proposed framework's ability to enforce security policies for distributed microservices management, thus harvesting the MSA characteristics to ensure IoT application security needs.

\end{abstract}
\section{Introduction}

Internet of Things (IoT) solutions are rapidly expanding across a large range of domains (i.e., healthcare, industrial, commercial, and agriculture) to extract valuable information from the data. International Data Corporation estimates a total generation of nearly 80 Billion zettabytes of data in 2025 from 55.7 billion connected devices \cite{IDC2021}. The vast amount and variety of data processed by IoT applications make the security of the IoT ecosystem one of the critical aspects \cite{al2022ai}. IoT applications handle data with different sensitivity levels, such as highly sensitive, personal, and mission-critical data (e.g., health records and security camera footage) and low-sensitive data like weather data. Hence, the processing of IoT data needs to be done within trust boundaries depending on data sensitivity levels \cite{xiong2018enhancing,varadharajan2016data,pahl2018architecture}. Owing to the highly resource-constrained nature of the IoT devices, computationally expensive data processing tasks are often offloaded to distributed computing resources with more resource availability (i.e., edge data centres, private data centres or cloud) \cite{al2022ai}. To optimally utilise these heterogeneous resources, the state-of-the-art research explores the federation of geo-distributed edge and cloud resources managed by multiple infrastructure providers (i.e., multi-domain edge-cloud) along with the distributed deployment of IoT applications across them \cite{farzin2022flex,faticanti2023locality}.

Thus, the current IoT landscape sees a rise in MicroService Architecture (MSA) as the preferred software architecture for IoT applications accompanied by federated edge and cloud computing for application deployment \cite{pallewatta2023placement}. Edge computing brings cloud-like processing, networking and storage capabilities closer to the network edge. With this, Edge computing not only improves QoS parameters like latency and throughput of IoT services but also enhances the security and privacy of IoT data. Processing data using edge resources closer to the data sources reduces the data exposure during transmission and provides higher flexibility to perform processing within trusted boundaries \cite{atieh2021next}. Meanwhile, MSA provides data segmentation based on sensitivity and well-defined boundaries among independently deployable and scalable microservices \cite{miller2021towards}. Such characteristics of MSA support the implementation of granular security controls at the microservices level with secured boundaries among microservices, which result in isolation among data types with varying sensitivity levels \cite{gitlab2022}. However, the deployment and composition of distributed microservices within multi-domain edge-cloud computing resources pose some novel security challenges due to geographical and administrative divisions among resources. We highlight these challenges using a motivating scenario in the following section.

\vspace{-3.0mm}

\subsection{Motivating Scenario}\label{sec:motivatingScenario}
\vspace{-2.0mm}
We consider a UAV Path Finding Application developed using MSA  (see Fig. \ref{fig:motivation_scenario}). The application is decomposed into microservices to achieve data segmentation, such that the data with varying sensitivity levels remains isolated for processing by distinct microservices. The IoT application is deployed across the geo-distributed edge and cloud resources managed by multiple resource providers, creating a \textit{multi-domain edge-cloud environment} characterised by geographical and administrative divisions, resulting in security scenarios that affect microservice placement, scaling, migration and composition-related decisions:

\begin{figure}
    \centering
    \includegraphics[width=0.9\linewidth]{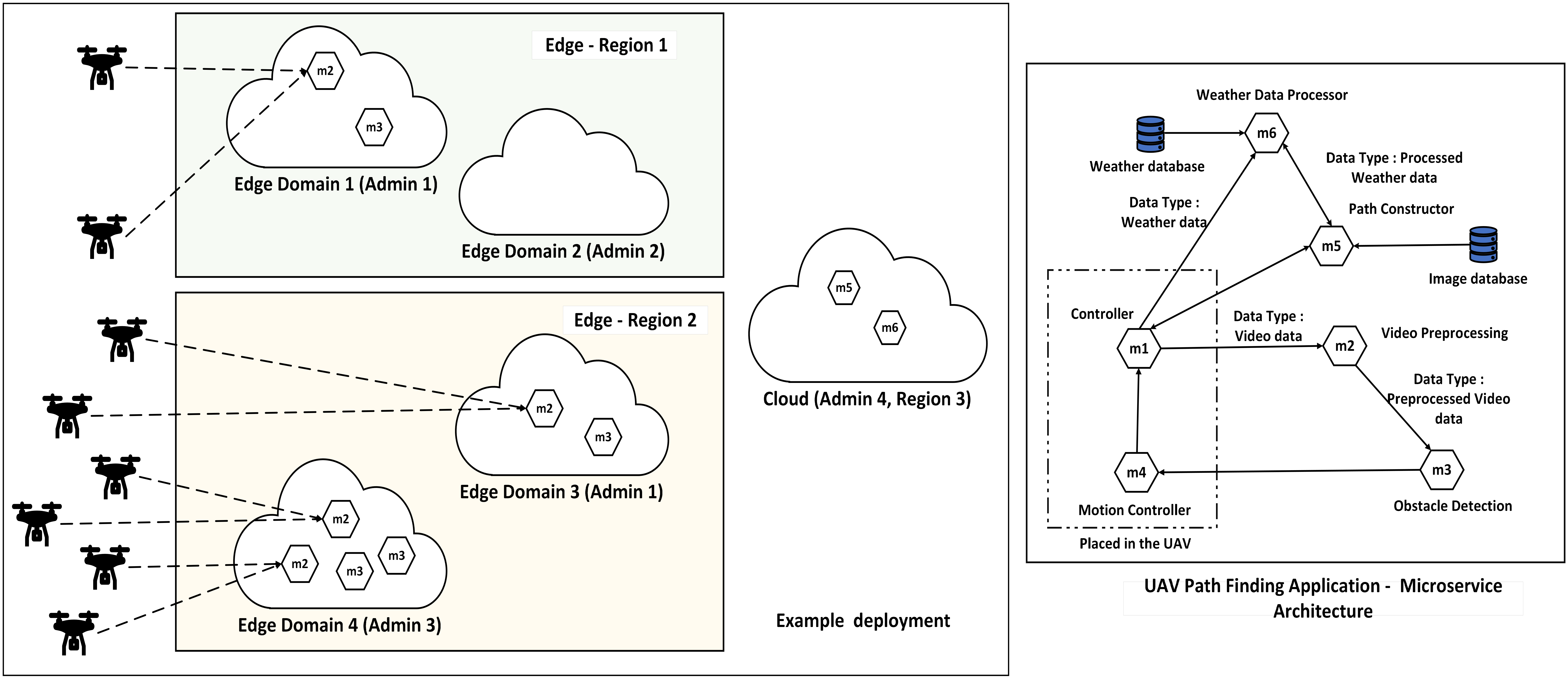}
    \vspace{-0.4cm}
    \caption{Motivating Scenario - Deployment Scenario of UAV Path Finding Application}
    \label{fig:motivation_scenario}
    \vspace{-0.7cm}
\end{figure}

    \textit{Domain-aware microservice placement restrictions -} Due to the high sensitivity of the live video data (i.e., footage containing private properties, critical infrastructure and individuals), placement of m2 and m3 are restricted to trusted Edge domains. In the example environment, this restricts the dynamic placement of m2 and m3 to domains managed by Admin 1 and Admin 3.  

      \textit{Locality constraints between the data source and ingress microservice -} To reduce the exposure of sensitive data during transmission, the real-time video data sent from UAVs need to be processed by m2 microservice instances placed within the same domain the UAV connects to. Thus, Edge Domain 4 should contain more microservice instances to satisfy the request volume generated from a comparatively higher number of UAVs connected to Edge Domain 4 to meet performance requirements without forwarding requests to m2 instances in other domains.

    \textit{Locality constraints on microservice composition -}  m2 pre-processes privacy-sensitive video data and forwards it to m3 for obstacle detection. Pre-processing actions like aggregation, anonymizing and transformation make the data output of m2 less privacy-sensitive than its input. However, as it still can contain some privacy-sensitive contextual data, it should be processed without leaving the region to reduce the exposure during data transmission between collaborating microservices (m2 and m3 in this case). In this example, resource-constrained Edge Domain 4 has insufficient resources to host m3 instances to support all its directly connected UAVs. Thus, more m3 instances are placed within Edge Domain 3 within the same region as Edge Domain 4 to satisfy the throughput requirements without violating the security policy.

The above scenarios demonstrate that microservices can have granular security constraints depending on the sensitivity of the data processed by/transmitted among microservices. Security constraints govern two types of microservice management activities: where the microservices are deployed during their initial placement, scaling and migration, and request routing among horizontally scaled microservices placed across domains. The following research questions must be addressed to ensure the security constraints are satisfied during above activities:

 1. How to define and integrate such granular security constraints into the microservice deployment decision process in the context of multi-domain deployments?  With the popularity of MSA as a cloud-native software architecture, DevSecOps practices promote the use of Policy-As-Code (PaC) to decouple security policies from the application code \cite{chandramouli2022implementation} to improve dynamic policy enforcement. However, existing research in multi-domain Edge-Cloud environments \cite{faticanti2020throughput,deng2021fogbus2,pallewatta2023placement} does not explore the use of PaC to define microservice deployment and composition-related security constraints and integrate PaC to control planes for microservice deployment decision making. 
 
 2. How to enforce the defined security constraints during dynamic composition and request routing among multi-domain microservices? Microservices utilize the cloud-native technology stack encompassing container technologies (e.g., Docker), container orchestration (e.g., Kubernetes) and service mesh (e.g., Istio) to facilitate their dynamic deployment and composition within distributed computing environments. Thus, the method of security policy enforcement is intertwined with these underlying technologies and needs to be dynamically handled by the intelligent control plane logic. Existing architectural frameworks for edge and cloud integrated environments \cite{ermolenko2021internet,deng2021fogbus2,pallewatta2023placement} mainly focus on proposing control planes to ensure QoS requirements (e.g., latency, throughput, etc.) and lack focus on dynamic security policy enforcement through intelligent configuration of underlying technologies.

\vspace{-4.0mm}

\subsection{Contributions}

\vspace{-2.0mm}

To overcome these limitations, we propose an architectural framework that facilitates the integration and enforcement of security policies during the deployment and composition of microservices across multi-domain edge and cloud resources.

\noindent \textbf{Main contributions }of our work are as follows:

1. Derive the requirements of a secure microservice management framework for multi-domain Edge-Cloud environments by analyzing a microservices-based IoT application scenario to resolve the above research questions.

 2. Design an architectural framework that integrates PaC into an intelligent control plane to satisfy the identified requirements.

 3. Implement a prototype system based on the proposed framework using open-source cloud-native technologies.

 4. Integrate a PaC-integrated heuristic placement algorithm to evaluate the proposed architectural framework for secure management of microservices. 

Our proposed framework can be used by IoT application developers to integrate PaC to define and enforce security requirements, thus extending DevSecOps practices to the edge. 

Edge-cloud infrastructure providers and researchers can use the architectural framework to implement platforms for secure management of microservices-based IoT applications.

\vspace{-4.00mm}
\section{Related Work}
\vspace{-3.00mm}

This section compares state-of-the-art edge application management frameworks and platforms with our proposed framework. 

Early edge-cloud frameworks like FogBus \cite{tuli2019fogbus} focused on application execution directly on physical resources or virtual machines. However, with the emergence of MSA and lightweight deployment technologies like Docker containers, Edge computing frameworks were developed for efficient deployment and orchestration of containers. FogPlan \cite{yousefpour2019fogplan}, Con-Pi \cite{mahmud2021pi}, and FogBus2 \cite{deng2021fogbus2} proposed frameworks for dynamic provisioning of containerised applications. Works such as Wang et al. \cite{wang2022container}, FogAltas \cite{fogatlas}, and Foggy \cite{santoro2017foggy} support the management of containers at a scale by integrating Kubernetes orchestrator. Sophos \cite{Sophos} and MicroFog \cite{microfog2023} frameworks integrated Istio service mesh with Kubernetes to provide service discovery and load balancing for the dynamic composition of microservices. All of these frameworks consider the placement of microservices across integrated edge and cloud environments. However, only MicroFog supports distributed deployment and compositions of microservices across multiple Kubernetes clusters forming a multi-domain edge-cloud environment. 

All above frameworks focus on deploying applications in a performance-aware manner. Frameworks such as FogBus2 \cite{deng2021fogbus2}, FogAltas \cite{fogatlas} and MicroFog \cite{microfog2023} allow users to implement placement algorithms to meet QoS requirements such as latency and throughput. Out of these frameworks, Fogbus \cite{tuli2019fogbus} utilises Blockchain technology to add security by ensuring data integrity during data transmission among edge computing nodes. As a data privacy and security measure, FogAtlas \cite{fogatlas} allows users to select which data to keep on-premise and which data to send to the cloud. Use of Istio service mesh in MicroFog \cite{microfog2023} and Sophos \cite{Sophos} enables mutual TLS authentication by default, which encrypts the communication among microservices. However, none of the above frameworks focuses on integrating security constraints related to the dynamic deployment of microservices or restrictions on their dynamic composition, especially considering multi-domain edge-cloud environments. 

Thus, in this paper, we propose a framework designed for the secure management of microservices within multi-domain edge-cloud environments using PaC. It is important to note that our proposed framework does not replace the data security measures achieved through Blockchain technology or data encryption discussed in existing frameworks. Instead, this framework complements these measures by introducing an intelligent control plane that conducts microservice management (including dynamic deployment and composition) in a security policy-aware manner within multi-domain edge-cloud environments.

\vspace{-2.00mm}

\section{Secure Microservice Management Framework}

\vspace{-2.00mm}

This section details our \textbf{\textit{"Secure Microservice Management Framework" }}, containing requirement elicitation of the frameworks, system model representing the business domain, architectural framework depicted using Modular Architecture Diagram and Interaction Diagram and a prototype based on the architectural framework.


\vspace{-4.0mm}
 
\subsection{Requirement Elicitation}\label{sec:Req_Elicitation}
We follow the \textit{"Scenario-based requirement elicitation method"} proposed by Haumer et al. \cite{haumer1998requirements} to construct generalized requirements of the framework utilizing the motivating scenario from Section \ref{sec:motivatingScenario}. Requirements aim to resolve the two research questions related to secure microservice management by utilizing PaC. Elicited framework requirements are as follows:

\vspace{1mm}
\noindent 1. Provide an approach to define microservice management-related security constraints and enable their integration into deployment and composition decision process: Establish a method to define different types of security constraints related to MSA such as \textit{Domain-aware microservice placement restrictions}, \textit{Locality constraints between IoT device and ingress microservice}, and \textit{Locality constraints on microservice composition} using PaC. 
Policy data and policy evaluation functions included in the definitions should be accessible through APIs to ensure the integration of security policies into edge-cloud control planes.

\vspace{1mm}
\noindent 2. Ensure the utilization of defined policies during microservice deployment within multi-domain resources: Policies should be considered during the initial placement of microservices upon the arrival of new application placement requests and their migration or scaling based on runtime system monitoring metrics (e.g., resource utilization, request rates, etc). Edge-cloud control planes should have access to policy definitions through APIs to retrieve eligible domains for each microservice and provide them as input to placement, scaling or migration algorithms to determine where to deploy microservices.

\vspace{1mm}
\noindent 3. Ensure that the request routing among microservices complies with the security policies: When deploying microservice instances based on the output of the placement, scaling or migration algorithms, the control plane should configure the underlying service mesh to ensure that traffic routing adheres to locality-related security policies such as \textit{Locality constraints between IoT device and ingress microservice} and \textit{Locality constraints on microservice composition}.

\vspace{1mm}
\noindent 4. Ensure scalable management of policies such that the distributed control plane can efficiently retrieve the policy evaluations or policy-related data across the multi-domain edge and cloud clusters.

\vspace{1mm}
\noindent 5. The framework must be designed for extensibility, enabling the extension of existing policies (e.g., new locality levels, new domain definitions, etc.) and integration of new security policy types.

\vspace{-4.00mm}

\subsection{Domain Model}\label{sec:domainModel}
\begin{figure*}[!ht]
    \includegraphics[width=\linewidth]{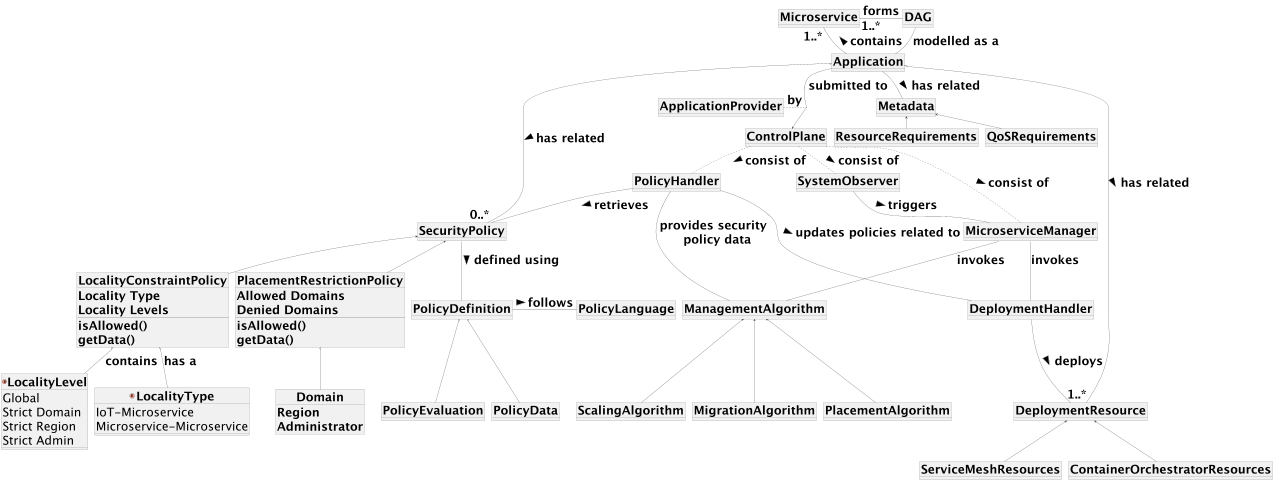}
    \vspace{-0.4cm}
    \caption{Domain Model for secure microservice management framework}
    \label{fig:domain_model}
    \vspace*{-0.7cm}
\end{figure*}

\vspace{-2.00mm}
Based on the requirements, we construct a domain model (see Fig. \ref{fig:domain_model}) to capture the requirements and depict the domain logic managed by our framework. 

\textit{IoT Application Providers} submit requests to the \textit{Control Plane} to conduct the initial placement of MSA-based \textit{Applications}. According to existing literature, MSA applications can be modelled as Directed Acyclic Graphs (DAGS) \cite{guerrero2019evaluation,faticanti2020throughput}, thus enabling the control plane to utilize \textit{DAG} application model, \textit{Deployment Resources} and \textit{Security Policies} along with other \textit{Metadata} such as \textit{QoS Requirements} and \textit{Resource Requirements} to manages microservices across distributed edge-cloud clusters to meet \textit{Security Policies} and \textit{QoS Requirements}.

\textit{Control Plane} contains three main components: \textit{Microservice Manager, System Observer and Policy Handler.} \textit{Microservice Manager} is responsible for invoking \textit{Management Algorithms} including: 1) \textit{Placement Algorithms} that are invoked during initial placement of the \textit{Application}, 2) \textit{Scaling Algorithm}s and \textit{Migration Algorithm} that are invoked based on triggers received from the \textit{System Observer} to optimize the deployment of already placed \textit{Applications}. \textit{System Observer} communicates with components that monitor performance metrics across edge-cloud clusters (i.e., Prometheus, Kubernetes Metric Server, etc.) and trigger \textit{Scaling Algorithms} or \textit{Migration Algorithms} accordingly. \textit{Policy Handler} is responsible for integrating \textit{Security Policy} related \textit{Policy Data} and \textit{Policy Evaluations} when executing \textit{Management Algorithms} to ensure that \textit{Security Policies} are enforced throughout the management cycle of microservices. 

\textit{Policy Handler} acts as the interfacing component that retrieves the data and evaluations from Security Policies defined using PaC. The Policy Definitions are created following a Policy Language and define Policy Data and Policy Evaluation functions. Domain model depicts three secure microservice management-related policy types derived from the motivation scenario as follows:

 1. Placement Restriction Policy: Defines the \textit{Domain-aware Microservice Placement Restrictions}, with Policy Data containing allowed/ denied domains for the microservices (exposed through getData function endpoint) and Policy Evaluation (exposed through isAllowed function endpoint) to evaluate and return the eligibility of the microservice to domain mappings.

 2. Locality Constraint Policy (Locality Type: IoT-Microservice): Defines the \textit{ Locality constraints between IoT Devices and Ingress Microservices}, where Policy Data contains the \textit{Locality Level}. This policy also has two endpoints exposing Policy Data (i.e., locality level for each ingress microservice) and Policy Evaluation calculating if the placement of an ingress microservice is eligible given the domain of the IoT device.

 3. Locality Constraint Policy (Locality Type: Microservice-Microservice): Defines the \textit{Locality constraints between communicating microservices}. This contains Policy Data denoting \textit{Locality Levels} between consumer and consumed microservices. Policy Evaluation calculates if the placement of a consumed microservice is allowed given the domain of the consumer microservice.

Microservice Manager uses the policy data and evaluations retrieved through the Policy Handler during two main steps:

1. During the execution of the Management Algorithms: 
Microservice Manager uses Security Policy data/evaluations queried through the Policy Handler as input to Placement Algorithms, Scaling Algorithms and Migration Algorithms to determine where to deploy each microservice in a security policy-aware manner.

 2. During microservice deployment using Deployment Resources: After receiving microservice to device mapping from the Management Algorithm, the Microservice Manager uses Security Policy data/evaluations queried through the Policy Handler to update Deployment Resources to enforce security policies (e.g.,  update routing rules among microservices by updating Virtual Services, Destination Rules of Istio service mesh). Afterwards, the Microservice Manager invokes the Deployment Handler to execute the updated Deployment Resources.

\vspace{-4.00mm}
\subsection{Framework Design} \label{sec:framework_design}
\vspace{-1.00mm}
Based on the domain model, this section designs the architectural framework for the secure management of microservices within multi-domain edge-cloud environments using two views: 1) a component diagram depicting the modular architecture and 2) an interaction diagram depicting the dynamic behaviour.

\vspace{-4.00mm}
\subsubsection{Component Diagram:} 
\begin{figure}[!h]
    \centering
    \includegraphics[width=0.85\linewidth]{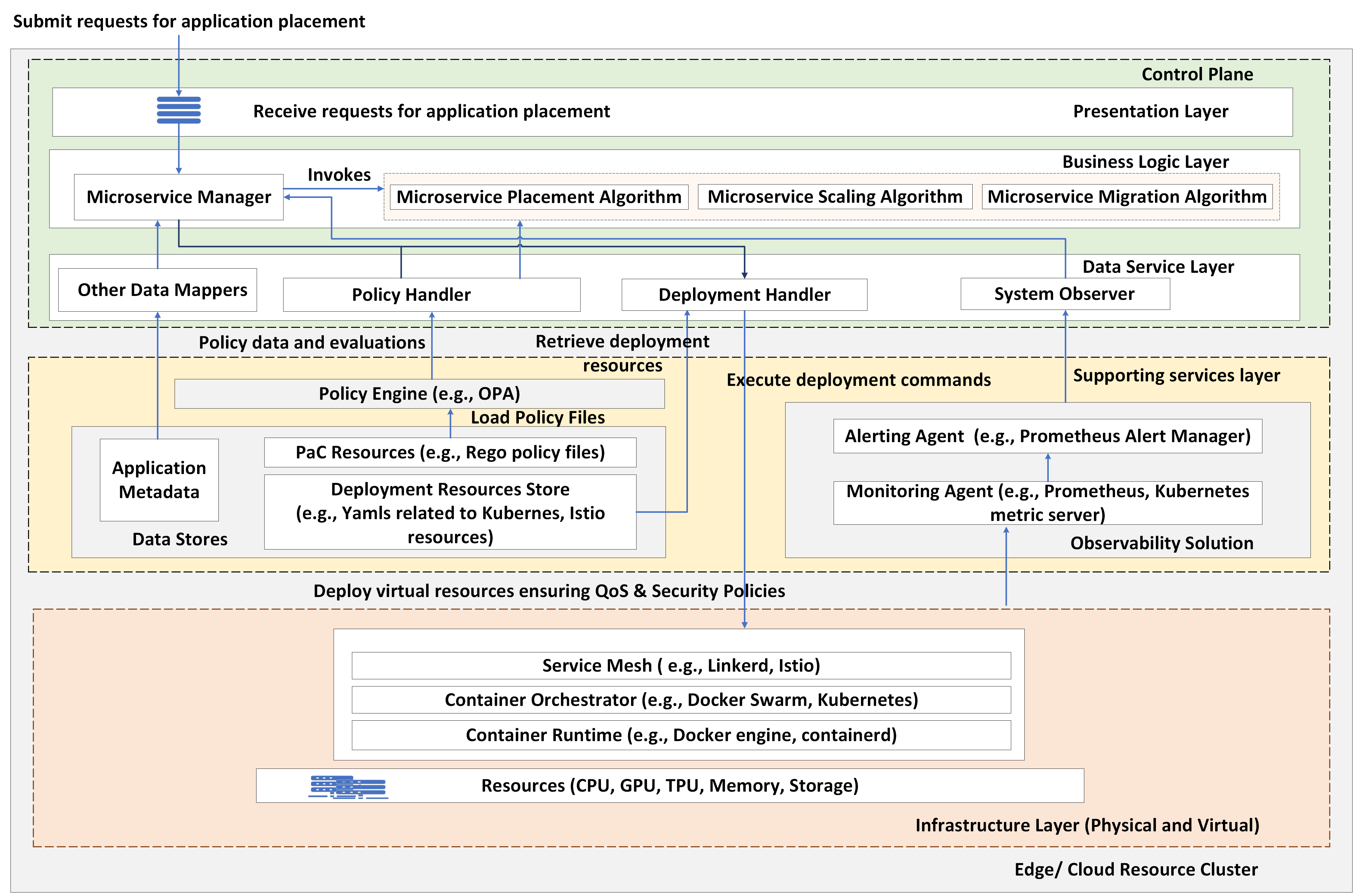}
    \caption{Component Diagram visualizing the highlevel architecture of the framework}
    \label{fig:component_diagram} 
    \vspace{-5.0mm}
\end{figure}

The architectural framework consists of three main layers: Infrastructure Layer, Supporting Services Layer and Control Plane (see Fig. \ref{fig:component_diagram}). 
\vspace{-5.0mm}

\textbf{Infrastructure Layer} provides the infrastructure-level support for running microservices as containers at a scale within distributed environments. This layer contains physical resources and virtual machines (e.g., CPU, GPU, TPU, Memory and Storage) as the underlying foundation with a Container Runtime (e.g., Docker engine \footnote{\url{https://www.docker.com/}}, containerd \footnote{\url{https://containerd.io/}}, etc.) for running containers. Following the cloud-native practices, a Container Orchestrator (e.g., Docker Swarm \footnote{\url{https://docs.docker.com/get-started/swarm-deploy/}}, Kubernetes \footnote{\url{https://kubernetes.io/}}, etc.) manages the containers at a scale and a Service Mesh (e.g., Istio \footnote{\url{https://istio.io/latest/}}, Linkerd \footnote{\url{https://linkerd.io/}}) enables dynamic composition of containerised microservices across domains.

\textbf{Supporting Service Layer} consists of containerised \textit{Observability Solution} and \textit{Data Stores}, providing services to the Control Plane. \textit{Data Stores} maintain data required by the Control Plane to make intelligent microservice management decisions. Data Stores include \textit{Application Metadata} describing the application, \textit{Deployment Resources}, which are the resource definitions used by Container Orchestrator and Service Mesh to deploy containerised microservices and \textit{PaC Resources}, which are the security policy definition files related to each application. \textit{Observability Solution} monitors the infrastructure layer through \textit{Monitoring Agents} and generates alerts using an \textit{Alerting Agent}.

\textbf{Control Plane} consists of three main layers as follows: 

\noindent 1. \textit{Presentation Layer} provides the users access to the Control Plane through a REST API or a GUI to submit the application.
    
\noindent 2. \textit{Business Logic Layer} contains intelligence related to secure microservice management encapsulated in microservice placement, scaling and migration algorithms. Microservice Manager invokes these algorithms on two occasions: 1) When a request for an application placement is received through the presentation layer, 2) When the System Observer communicates the need for microservice scaling or migration. The Microservice Manager also implements the logic to update Deployment Resources according to the algorithm output and communicates with the Deployment Handler to retrieve and deploy the Deployment Resources. To enable the secure management of microservices, Microservice Manager and Management Algorithms obtain security policy-related data using the Policy Handler component in the Data Service Layer.
    
\noindent 3. \textit{Data Service Layer} is responsible for communicating with Supporting Services layer and Infrastructure layer. It acts as the abstract layer between the Business Logic Layer and lower layers, and consists of 4 main components: 1) Policy Handler responsible for loading policy files to the Policy Agent and querying the Policy Agent for security policies (i.e., policy data and evaluation results), 2) Deployment Handler responsible for retrieving deployment resource definitions from the Supporting Services Layer, updating them based on the input from the Microservice Manager and executing them on the Infrastructure Layer, 3) System Observer responsible for receiving alerts from Observability Solution and directing the relevant alerts to the Microservice Manager, 4) Other Data Mappers responsible for receiving Metadata used as input for management algorithms.

\vspace{-3.00mm}

\subsubsection{Interaction Diagram}

\begin{figure*}[!ht]
    \includegraphics[width=0.95\linewidth]{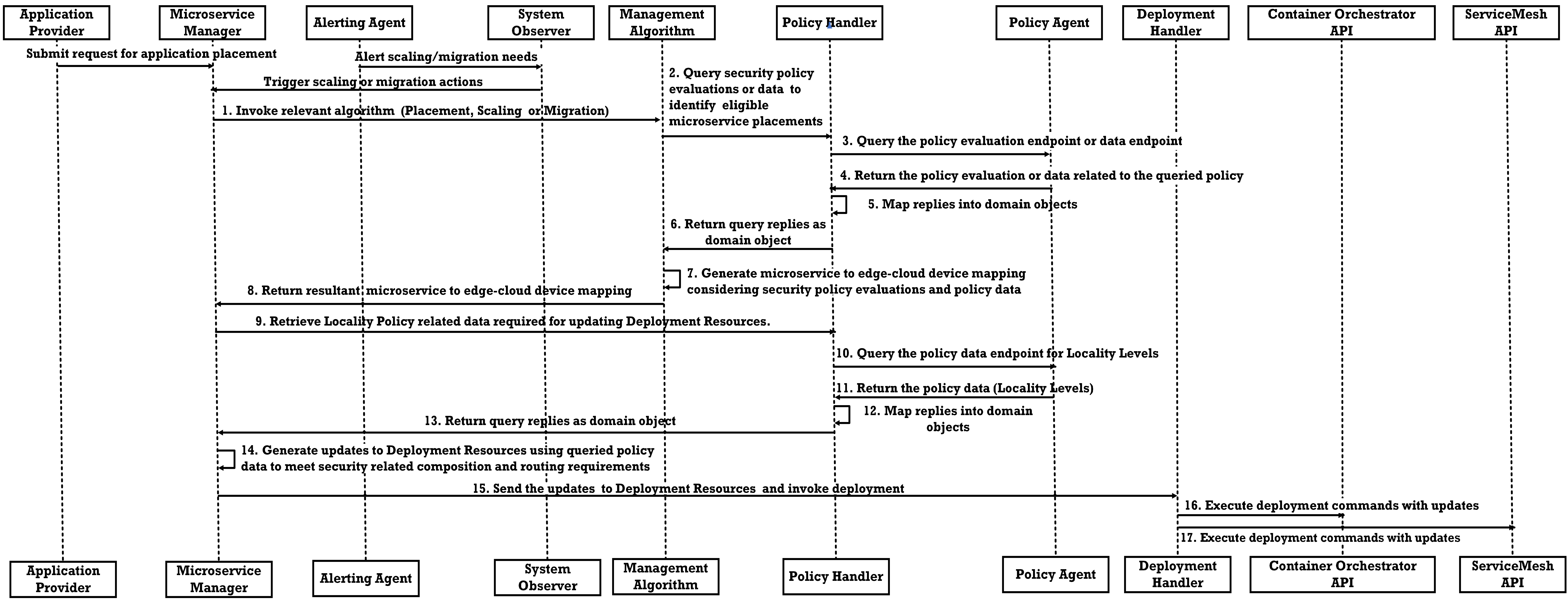}
    \caption{Interaction Diagram}
    \label{fig:sequenceDiagram}
    \vspace*{-0.6cm}
\end{figure*}

Figure \ref{fig:sequenceDiagram} depicts the ordered interactions among framework components, demonstrating the incorporation of security policies into the decision process.

\vspace*{-2.0mm}

\subsection{Prototype Implementation}
To demonstrate and evaluate the viability of the framework, we create a prototype by extending MicroFog \cite{microfog2023}: an open-source platform containing a base architecture for distributed placement of microservices across multiple edge-cloud domains. MicroFog contains a distributed controller that communicates with Kubernetes (as Container Orchestrator) and Istio (as Service Mesh) to automatically deploy microservices as Docker containers (as Container Runtime). However, MicroFog focuses on satisfying QoS parameters during the initial placement and does not consider application security requirements. We utilised and extended Microfog components to implement the Supporting Service Layer and Control Plane of our architectural framework to enable secure management of microservices as follows:

\noindent \textbf{Supporting Services Layer Implementation - } 

    \noindent\textit{Data Stores:} MicroFog uses Redis \footnote{\url{https://redis.io/}} to store Application Metadata and MinIO \footnote{\url{https://min.io/}} to store Yaml files of the Deployment Resources. To integrate PaC, we extend MinIO to store the PaC Resources (i.e., Policy-as-Code files). The deployment architecture of MinIO in MicroFog contains distributed deployment, replication and fault-tolerance, which matches the requirement of scalable management of policies in distributed environments. We select Rego \footnote{\url{https://www.openpolicyagent.org/docs/latest/}}, a declarative language supporting fine-grained policy declaration to define the policies.

    \noindent \textit{Policy Agent:} We integrate Open Policy Agent (OPA) \footnote{https://www.openpolicyagent.org/}, a cloud-native policy management tool which reads Rego policies and provides HTTP API endpoints for accessing policy data and evaluations. We deploy OPA as Kubernetes pods and expose it to Control Plane as Kubernetes and Istio services.


    \noindent \textit{Observability Solution:} Kubernetes Metric Server as the Monitoring Agent and a custom microservice as the Alerting Agent to retrieve metrics, generate alerts and communicate them to the System Observer.

    \noindent \textbf{Control Plane Implementation -} We extend the \textit{Control Engine} component of MicroFog to support the architecture of our proposed \textit{Control Plane}. MicroFog \textit{Control Engine} contains a data mapper Java object to access Application Metadata and a Deployment Handler object to communicate with the MinIO bucket to load, update and deploy Kubernetes and Istio resources. We utilize these and extend \textit{Control Engine} by implementing Policy Handler and System Observer components to complete our proposed Data Service Layer. 
    
    \noindent \textit{Policy Handler}: We implement the \textit{Policy Handler} as a collection of Java objects that provides methods for querying the policy data and evaluations. Policy Handler uses a Reactive REST Client to send \textit{asynchronous requests} to OPA as parallel queries to improve the performance during policy querying. Moreover, Policy Handler maps the replies received from the OPA into domain objects easily accessible to \textit{Business Logic Layer}.

    \noindent \textit{System Observer}: System Observer contains a REST API endpoint to receive alerts from Alerting Agent. Upon receiving the alerts, System Observer directs them to the Microservice Manager to invoke scaling and migration actions.

    \noindent \textit{Microservice Manager and Management Algorithms}: 
    We extend MicroFog to consider security policies during deployment algorithm execution and deployment resource updates. Microservice Manager instantiates \textit{Policy Handler} component and loads Rego policies into the OPA for use during business logic execution. We implement the "PaC Integrated Microservice Placement Algorithm" (see Algorithm \ref{algorithm}): a heuristic Placement Algorithm invoked by the Microservice Manager. The algorithm queries the loaded security policies through \textit{Policy Handler} to identify eligible placement for microservices. Furthermore, we extend the Microservice Manager to utilize locality-related security policies to update Istio Virtual Services and Destination Rules to ensure that request routing among microservices adheres to locality levels among them. 
    
    \vspace{-0.7cm}
    \begin{algorithm}[!h]
		\caption{PaC Integrated Microservice Placement}\label{algorithm}
		\begin{algorithmic}
                \algorithmicsize
			\Procedure{PlaceApplication}{}
			\State \textit{application} $\gets$ received by Control Plane for placement
   
                \State {\textbf{\# Microservices that are placed within IoT devices}} 
                
			\State \textit{placedMicroservices} $\gets$  placed on IoT devices

                \State{\textbf{\# Microservices that directly receives data/ requests from IoT devices  }}
                
			\State \textit{ingressMicroservices} $\gets$ get ingress microservices of application 

                  \State {\textbf{\# Start from ingress microservices and traverse through the application graph iteratively}} 
			\State $microservicesToPlace$ $\gets$ get eligible microservices for placement based on \textit{placedMicroservices} and \textit{ingressMicroservices}

                \State {\textbf{\# Sort based on the strictness of the locality constraints to ensure microservices with stringent locality constraints are prioritised}} 
                
                \State $sortedMicroservicesToPlace$ $\gets$ sort microservicesToPlace by locality constraints

                \While{ $sortedMicroservicesToPlace$}

                 \State $microservice$ $\gets$ first in $sortedMicroservicesToPlace$

                 \State $eligibleDomains$ $\gets$ get eligible domains based on \textbf{\textit{Placement Restriction Policy}}

                 \If{microservice is an ingress microservice}
                    \State $eligibleDomains$  $\gets$ refine $eligibleDomains$ based on \textbf{\textit{Locality Constraint Policy (IoT-Microservice)}}
                \Else 
                    \State $eligibleDomains$  $\gets$ refine $eligibleDomains$ based on \textbf{\textit{Locality Constraint Policy (Microservice-Microservice)}}
                \EndIf 

                \State {\textbf{\# Device to microservice mapping  used for deploying microservices to meet throughput expectation of the application}} 
                
                 \State $devicesToPlace$ $\gets$ select devices with sufficient resources from $eligibleDomains$

                 \State Update $placementMapping$

                 \State Update $placedMicroservices$

                 \State  $microservicesToPlace$ $\gets$ update eligible microservices for placement based on \textit{placedMicroservices} and \textit{ingressMicroservices}

                 \State $sortedMicroservicesToPlace$ $\gets$ sort microservicesToPlace by locality constraints

               \EndWhile

               \State \textbf{return } $placementMapping$



			\EndProcedure
		\end{algorithmic}
	\end{algorithm}

    \vspace{-0.8cm}

    \noindent \textit{Presentation Layer:} MicroFog provides a REST API to submit application placement requests. We extend this API, to support multi-domain application placement by enabling the user to include details about expected request distribution across domains (i.e., throughput distribution).

\vspace{-4.0mm}
\section{Evaluation}
We demonstrate the  prototype's ability to meet the requirements identified in Section \ref{sec:Req_Elicitation} which answers the two research questions explored in this work.

\vspace{-3.0mm}

\subsection{Experiment Setup} 

\noindent \textit{Multi-domain Edge-Cloud Infrastructure -} Following the motivation scenario, we consider two edge computing clusters (representing the Region 2 in Fig. \ref{fig:motivation_scenario} with Edge Domain 3 and Edge Domain 4) and one cloud cluster (representing the Region 3 in Fig. \ref{fig:motivation_scenario}). They  are created as KinD \footnote{https://kind.sigs.k8s.io/} Kubernetes
(containerised k8s) clusters communicating using Istio service mesh in multi-primary mode.

\noindent {\textit{Deployed application -} We create the UAV Path finding application from section \ref{sec:motivatingScenario} using the Workflow Generator from \cite{microfog2023}. The security policy definitions are derived from the 3 example scenarios (section \ref{sec:motivatingScenario}) and mapped onto the 3 policy types defined in Domain Model (see Fig. \ref{fig:regoPolicies})

\vspace{-0.4cm}
\begin{figure*}[!ht]
    \includegraphics[width=\linewidth]{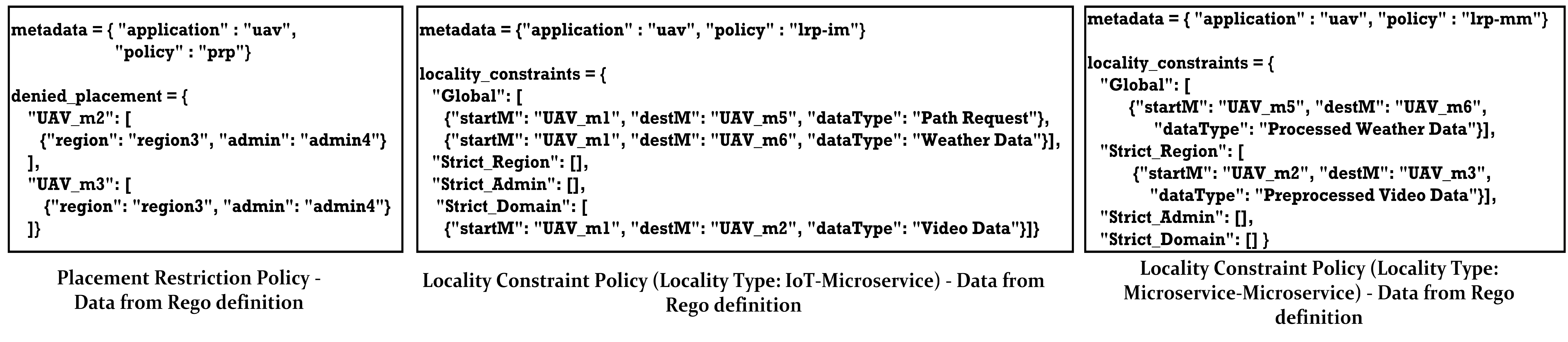}
    \vspace{-0.7cm}
    \caption{UAV Path Finding Application - Security Policies}
    \label{fig:regoPolicies}
    \vspace{-5.0mm}
\end{figure*}

\noindent {\textit{Microservice Management Scenario - }Application providers of the UAV Path Finding Application submit a request to deploy the application across \textit{Multi-domain Edge-Cloud Infrastructure}. The request is submitted with the expected throughput per edge domain. We consider a scenario similar to Fig. \ref{sec:motivatingScenario} where Edge Domain 4 expects twice the throughput as Edge Domain 3. Afterwards,  Microservice Manager in the \textit{Control Plane} guides the deployment of the microservices with updated YAML resources to ensure microservice placement and dynamic request routing satisfy the security policies.

\vspace{-2.00mm}

 \subsection{Analysis of Results}


After the control plane completes the application deployment, we evaluate the framework based on the two main management actions performed by the Microservice Manager to resolve our highlighted research questions: 1) microservice placement should satisfy the security policies while meeting the throughput requirements of the application. 2) dynamic request routing among distributed microservices across domains should satisfy the related security policies.

\begin{figure}[h]
    \centering
    \includegraphics[width=0.85\linewidth]{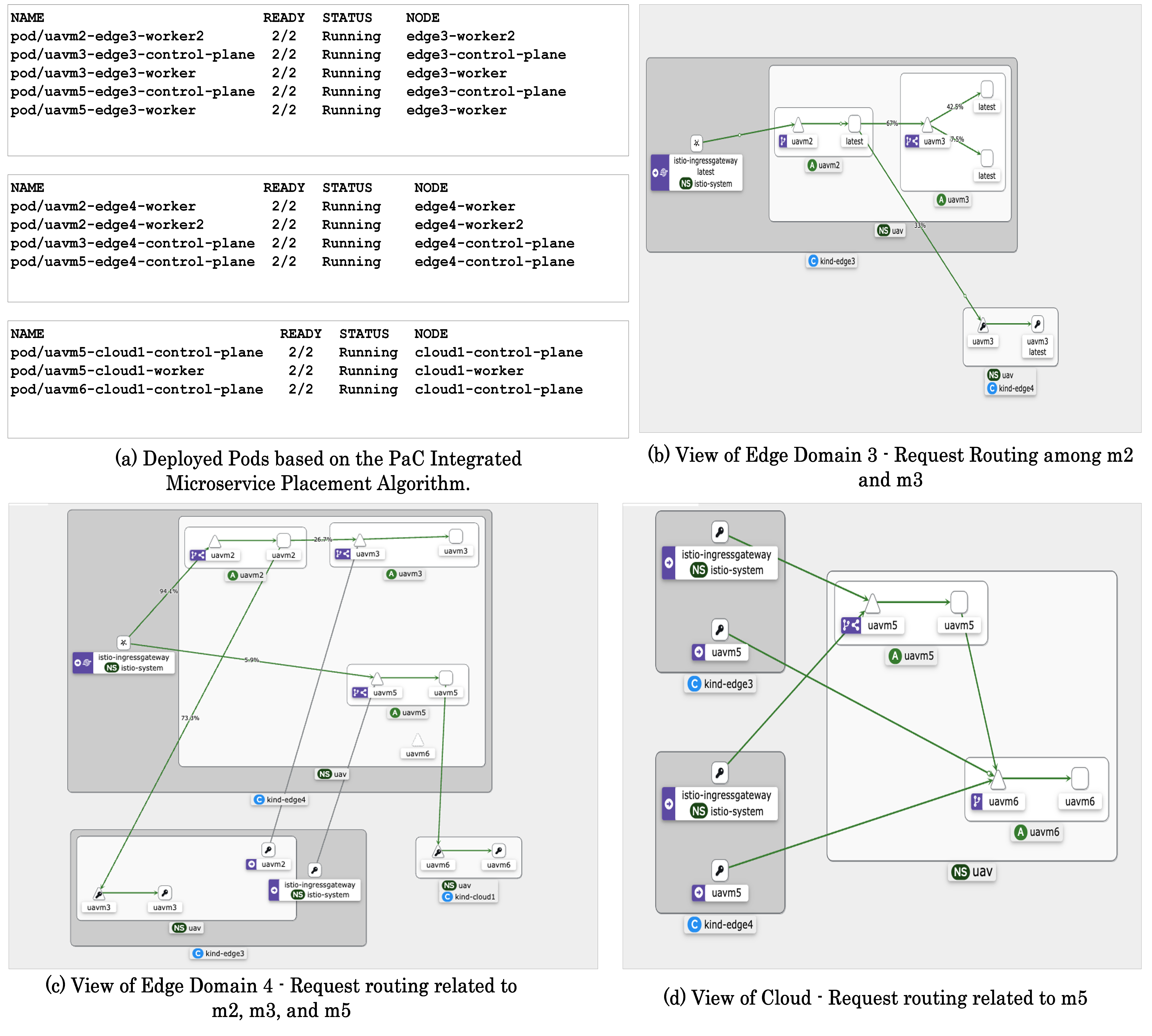}
    \vspace{-0.4cm}
    \caption{Deployed pods and their composition across domains}
    \label{fig:pods_deployment}
    \vspace{-0.4cm}
\end{figure}

Fig. \ref{fig:pods_deployment} (a) contains the Kubernetes API output listing the pods deployed within the UAV application namespace in each edge/cloud domain. Deployed locations of each microservice confirm that the \textit{placementMapping} generated from the \textit{Control Plane} satisfies the security policies defined in Fig. \ref{fig:regoPolicies}. Resultant placement satisfies the Placement Restriction Policy by deploying microservices \textit{m2} and \textit{m3} within Edge Domain 3, and Edge Domain 4. Edge Domain 4 has twice the amount of \textit{m2} instances as Edge Domain 3 to meet the throughput requirement without violating the Locality Constraint Policy (Locality Type: IoT-Microservice). According to Locality Constraint Policy (Locality Type: Microservice-Microservice), \textit{m3} is not limited strictly to the domain of \textit{m2}  and can be placed within the region. Thus, Edge Domain 3 and Edge Domain 4 host \textit{m3}  to collectively satisfy requests generated from \textit{m2} instances (see Fig. \ref{fig:pods_deployment} (b)). Hence, the pod deployment demonstrates the ability of the Control Plane to integrate PaC and intelligently utilize them to make secure microservice deployment decisions while satisfying QoS parameters like throughput.

\begin{figure}[h]
    \centering
    \includegraphics[width=0.85\linewidth]{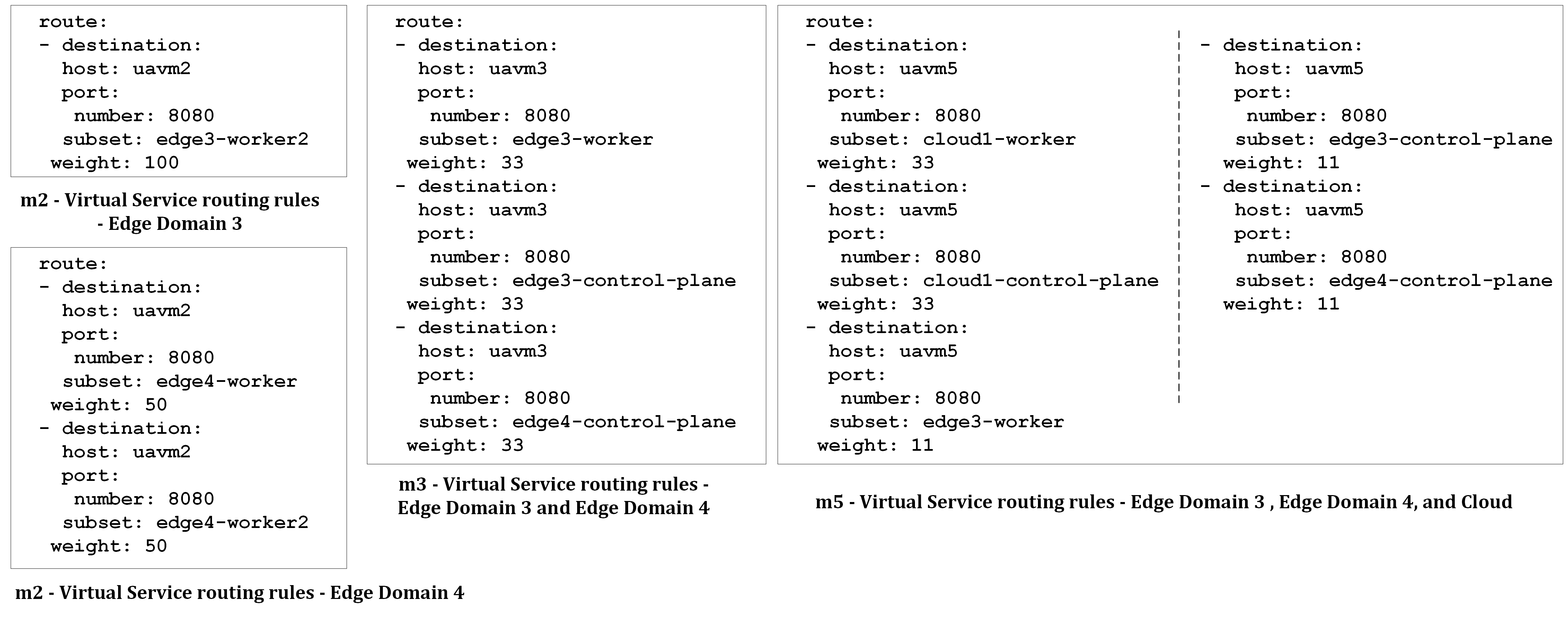}
    \vspace{-0.4cm}
    \caption{Routing rules in Istio Virtual Services updated by Control Plane}
    \label{fig:virtual_services}
    \vspace{-0.4cm}
\end{figure}

Istio Virtual Service definitions and Kiali views of the inter-microservice traffic flow demonstrate the dynamic composition and request routing among microservices. Figure \ref{fig:virtual_services} contains the routing rules of the Istio Virtual Services deployed within each domain. In our proposed framework, the \textit{Microservice Manager} generates these rules dynamically based on the Placement Mapping and Security policies retrieved from PaC. Afterwards, the Deployment Handler updates the Virtual Services with generated routing rules and deploys them in relevant domains. Figure \ref{fig:virtual_services} shows that the routing rules for \textit{m2} differ based on the domain to ensure that the requests received for \textit{m2} do not leave the domain and are served by \textit{m2} instances in the same domain (i.e., ensuring Strict Domain Locality level). Compared to this,  \textit{m3} routing rules are the same for both Edge Domains 3 and 4, enabling load balancing among \textit{m3} within the region (i.e., ensuring Strict Region Locality level). Due to the Global locality level, \textit{m5} routing rules allow request routing among regions, covering both edge and cloud clusters.

Kiali views related to traffic routing further demonstrate the impact and validity of the updated routing rules. As Kiali cannot capture all traffic routing in multi-cluster scenarios with a single view, we present multiple views covering all 3 domains (see Fig. \ref{fig:pods_deployment} - b,c,d). Views of Edge Domain 3 and Edge Domain 4 confirm that inter-domain request routing among edge clusters of the same region occurs for requests from \textit{m2} to \textit{m3}. Meanwhile, requests for \textit{m2}, which are received through the Ingress Gateway of the domain, do not cross domain boundaries. View of the Cloud demonstrates that \textit{m5} receives requests routed across regions which comply with the Global locality level.

Above results demonstrates framework's ability to meet requirements 1-3 of section \ref{sec:Req_Elicitation}.

\vspace{-5.00mm}
\subsection{Discussion on Extensibility and Scalability of Policy Handling (requirements 4-5 from section \ref{sec:Req_Elicitation}) }
\vspace{-0.2cm}
The Policy Handler in the Data Service Layer abstracts the policy querying process from the upper Business Logic Layer. Thus, the Policy Handler is responsible for converting domain objects to queries and replies from the Policy Agent to domain objects. Such layered architecture with separation of responsibility enables easy extension of existing security policies and adding new policy types. The use of a Policy Engine in the Supporting Services Layer ensures the scalable management and querying of the security policies. Being an integral part of the cloud ecosystem, Policy Engines such as OPA are designed with scalability in mind. Policy Engines have optimised query performance with concurrent query support and distributed deployment support to scale with the request load. 

\vspace{-5.00mm}

\subsection{Discussion}

\vspace{-2.00mm}

Results demonstrate the ability of the framework to successfully integrate PaC-based application security definitions to an intelligent \textit{Control Plane} for secure management of microservices within a multi-domain edge-cloud environment. PaC enables the flexibility for the \textit{Control Plane} to retrieve and utilize policy data/ evaluations during different stages of the microservice management process (e.g., during the execution of \textit{Microservice Management Algorithms} and updating Yaml resources related to microservice composition). The framework also demonstrated the ability to successfully handle multiple policy definition types with the use of \textit{Policy Handler} component that abstracts the communication among Control Plane and \textit{Policy Agent}. These observations and results solidify our framework's role as a foundational architectural blueprint for facilitating the development of platforms for secure microservice management, with inherent extensibility to enhance the Control Plane to handle edge-cloud use cases with diverse security requirements.

\vspace{-4.00mm}

\section{Conclusions and Future Work}
\vspace{-2.00mm}

In this paper, we present a framework for secure management of microservices during their deployment and dynamic composition within multi-domain edge-cloud environments. Our proposed framework uses PaC to define and manage granular microservice security policies. Policies are integrated into an intelligent control plane that utilises them to ensure security requirements. We created a prototype of the framework and evaluated its ability to achieve secure microservice management. The results validate the functionality of the framework, demonstrating its potential as a foundational architecture for facilitating the development of platforms for secure management of distributed microservices. In future work, we plan to improve the framework to handle security policy updates by integrating this framework into CI/CD pipelines.
\vspace{0.4cm}
\begin{credits}

\noindent \textbf{Data Availability Statement }As this project is funded by an industry partner, we are unable to publish the source code at this stage. To increase reproducibility, we extended an open-source framework along with open-source tools and explained the implementation of our framework in detail. High quality images of all figures used in the manuscript are available at \textcolor{blue}{\url{https://doi.org/10.5281/zenodo.12524961}}.

\subsubsection{\ackname} This work has been supported by the Cyber Security Cooperative Research Centre Limited whose activities are partially funded by the Australian Government’s Cooperative Research Centre Program.

\end{credits}

\vspace{-0.3cm}
\bibliographystyle{splncs04}
\bibliography{./bibliography/bibliography.bib}

\end{document}